\def\year{2022}\relax
\documentclass[letterpaper]{article} 
\usepackage{aaai22}  
\usepackage{times}  
\usepackage{helvet}  
\usepackage{courier}  
\usepackage[hyphens]{url}  
\usepackage{graphicx} 
\def\UrlFont{\rm}  
\usepackage{natbib}  
\usepackage{caption} 
\DeclareCaptionStyle{ruled}{labelfont=normalfont,labelsep=colon,strut=off} 
\usepackage{algorithmic}
\usepackage[vlined,ruled,linesnumbered]{algorithm2e}

\usepackage{subcaption}
\usepackage{amsmath}
\usepackage{braket}

\usepackage{newfloat}
\usepackage{listings}

\usepackage{xtab}
\usepackage{longtable}

\title{Qubit Routing using Graph Neural Network aided Monte Carlo Tree Search}

\author{
    Animesh Sinha$^*$,
    Utkarsh Azad$^*$,
    Harjinder Singh
}
\affiliations{
    Center for Computational Natural Sciences and Bioinformatics, International Institute of Information Technology, Hyderabad.
    Center for Quantum Science and Technology, International Institute of Information Technology, Hyderabad.
}

\begin{document}

\maketitle

\begin{abstract}
Near-term quantum hardware can support two-qubit operations only on the qubits that can interact with each other. Therefore, to execute an arbitrary quantum circuit on the hardware, compilers have to first perform the task of qubit routing, i.e., to transform the quantum circuit either by inserting additional SWAP gates or by reversing existing CNOT gates to satisfy the connectivity constraints of the target topology.  The depth of the transformed quantum circuits is minimized by utilizing the Monte Carlo tree search (MCTS) to perform qubit routing by making it both construct each action and search over the space of all actions. It is aided in performing these tasks by a Graph neural network that evaluates the value function and action probabilities for each state. Along with this, we propose a new method of adding mutex-lock like variables in our state representation which helps factor in the parallelization of the scheduled operations, thereby pruning the depth of the output circuit. Overall, our procedure (referred to as QRoute) performs qubit routing in a hardware agnostic manner, and it outperforms other available qubit routing implementations on various circuit benchmarks.
\end{abstract}

\maketitle


\section{\label{sec:intro}Introduction}

\begin{figure*}[t]
    \centering
    \includegraphics[width=0.9\textwidth]{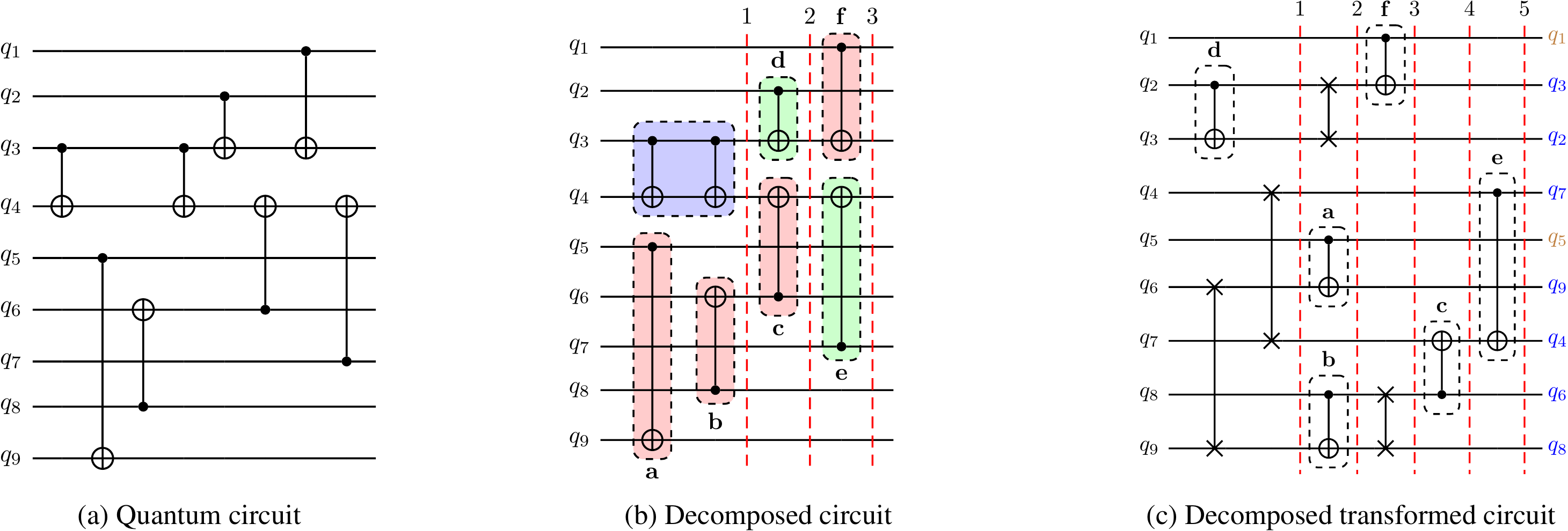}
    \caption{An example of qubit routing on a quantum circuit for 3$\times$3 grid architecture (Figure \ref{fig:topology-example}a). (a) For simplicity, the original quantum circuit consists only of two-qubit gate operations. (b) Decomposition of the original quantum circuit into series of slices such that all the instructions present in a slice can be executed in parallel. The two-qubit gate operations: $\{d,e\}$ (green) comply with the topology of the grid architecture whereas the operations: $\{a, b, c, f\}$ (red) do not comply with the topology (and therefore cannot be performed). Note that the successive two-qubit gate operations on $q_3\rightarrow q_4$ (blue) are redundant and are not considered while routing. (c) Decomposition of the transformed quantum circuit we get after qubit routing. Four additional SWAP gates are added that increased the circuit depth to $5$, i.e., an overhead circuit depth of $2$. The final qubit labels are represented at the end right side of the circuit. The qubits that are not moved (or swapped) are shown in brown ($\{q_1, q_5\}$), while the rest of them are shown in blue.}
    \label{fig:routing-example}
\end{figure*}

The present-day quantum computers, more generally known as Noisy Intermediate-Scale quantum (NISQ) devices \citep{nisq_preskill} come in a variety of hardware architectures \cite{IBMQ, hardware_sycamore, hardware_rigetti_aspen, hardware_xanadu}, but there exist a few problems pervading across all of them. These problems constitute the poor quality of qubits, limited connectivity between qubits, and the absence of error-correction for noise-induced errors encountered during the execution of gate operations. These place a considerable restriction on the number of instructions that can be executed to perform useful quantum computation \cite{nisq_preskill}. Collectively these instructions can be realized as a sequential series of one or two-qubit gates that can be visualized more easily as a quantum circuit as shown in Fig. \ref{fig:routing-example}a \citep{others_childs}.

To execute an arbitrarily given quantum circuit on the target quantum hardware, a compiler routine must transform it to satisfy the connectivity constraints of the topology of the hardware \citep{qroute_tket}. These transformations usually include the addition of SWAP gates and the reversal of existing CNOT gates. This ensures that any non-local quantum operations are performed only between the qubits that are nearest-neighbors. This process of circuit transformation by a compiler routine for the target hardware is known as qubit routing \citep{qroute_tket}. The output instructions in the transformed quantum circuit should follow the connectivity constraints and essentially result in the same overall unitary evolution as the original circuit \citep{qroute_dqn2}.

In the context of NISQ hardware, this procedure is of extreme importance as the transformed circuit will, in general, have higher depth due to the insertion of extra SWAP gates. This overhead in the circuit depth becomes more prominent due to the high decoherence rates of the qubits and it becomes essential to find the most optimal and efficient strategy to minimize it \citep{qroute_tket, qroute_dqn1, qroute_dqn2}. In this article, we present a procedure that we refer to as \textit{QRoute}. We use Monte Carlo tree search (MCTS), which is a look-ahead search algorithm for finding optimal decisions in the decision space guided by a heuristic evaluation function \citep{mcts_bandit_1, mcts_bandit_2, mcts_uct}. We use it for explicitly searching the decision space for depth minimization and as a stable and performant machine learning setting. It is aided by a Graph neural network (GNN) \citep{nn_edge_conv}, with an architecture that is used to learn and evaluate the heuristic function that will help guide the MCTS.


\begin{figure}[tp]
    \centering
    \includegraphics[width=0.95\linewidth]{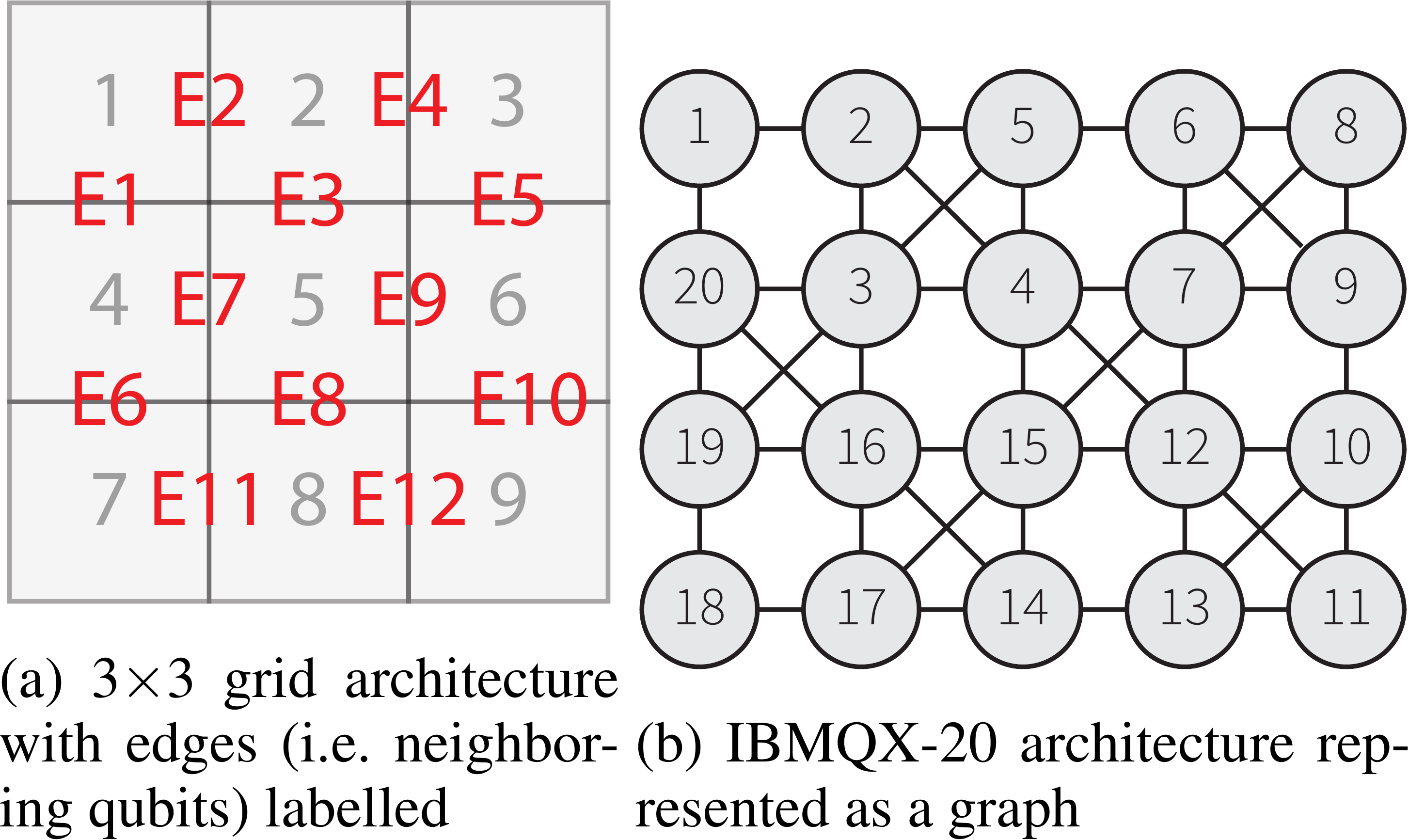}
    \caption{Examples of qubit connectivity graphs for some common quantum architectures}
    \label{fig:topology-example}
\end{figure}

\section{\label{sec:qubit-routing}Qubit Routing}
In this section, we begin by defining the problem of qubit routing formally and discussing the work done previously in the field.

\subsection{\label{sec:intro-defn}Describing the Problem}
The topology of quantum hardware can be visualized as a qubit connectivity graph (Fig. \ref{fig:topology-example}). Each node in this graph would correspond to a physical qubit which in turn might correspond to a logical qubit. The quantum instruction set, which is also referred to as quantum circuit (Fig. \ref{fig:routing-example}a), is a sequential series of single-qubit and two-qubit gate operations that act on the logical qubits. The two-qubit gates such as CNOT can only be performed between two logical qubits iff there exists an edge between the nodes that correspond to the physical qubits, \citep{qroute_dqn1}. This edge could be either unidirectional or bidirectional, i.e., CNOT can be performed either in one direction or in both directions. In this work, we consider only the bidirectional case, while noting that the direction of a CNOT gate can be reversed by sandwiching it between a pair of Hadamard gates acting on both control and target qubits \cite{utk_equiv_circuits}. 

Given a target hardware topology $\mathcal{D}$ and a quantum circuit $\mathcal{C}$, the task of qubit routing refers to transforming this quantum circuit by adding a series of SWAP gates such that all its gate operations then satisfy the connectivity constraints of the target topology (Fig. \ref{fig:routing-example}c). Formally, for a routing algorithm $\textrm{R}$, we can represent this process as follows:
\begin{equation}
\textrm{R}(\mathcal{C},\ \mathcal{D}) \rightarrow \mathcal{C}^{\prime}
\end{equation}
Depth of $\mathcal{C}^{\prime}$ (transformed quantum circuit) will, in general, be more than that of the original circuit due to the insertion of additional SWAP gates. This comes from the definition of the term \textit{depth} in the context of quantum circuits. This can be understood by decomposing a quantum circuit into series of individual slices, each of which contains a group of gate operations that have no overlapping qubits, i.e., all the instructions present in a slice can be executed in parallel (Fig. \ref{fig:routing-example}b). The depth of the quantum circuit then refers to the minimum number of such slices the circuit can be decomposed into, i.e., the minimum amount of parallel executions needed to execute the circuit. The goal is to minimize the overhead depth of the transformed circuit with respect to the original circuit.

This goal involves solving two subsequent problems of (i) qubit allocation, which refers to the mapping of program qubits to logic qubits, and (ii) qubit movement, which refers to routing qubits between different locations such that interaction can be made possible \citep{utk_qubit_noise}. In this work, we focus on the latter problem of qubit movement only and refer to it as qubit routing. However, it should be  noted that qubit allocation is also an important problem and it can play an important role in minimizing the effort needed to perform qubit movement.

\subsection{\label{sec:intro-related}Related Work}

The first major attraction for solving the qubit routing problem was the competition organized by IBM in 2018 to find the best routing algorithm. This competition was won by \citet{zulehner2018mapping}, for developing a Computer Aided Design-based (CAD) routing strategy. Since then, this problem has been presented in many different ways. These include graph-based architecture-agnostic solution by \citet{qroute_tket}, showing equivalence to the travelling salesman problem by \citet{paler_torus}, machine learning based methods by \citet{paler_ml}, and heuristic approaches by \citet{review-1}, \citet{review-2}, \citet{review-3}, etc. A reinforcement learning in a combinatorial action space solution was proposed by \citet{qroute_dqn1}, which suggested used simulated annealing to search through the combinatorial action space, aided by a Feed-Forward neural network to judge the long-term expected depth. This was further extended to use Double Deep Q-learning and prioritized experience replay by \citet{qroute_dqn2}. 

Recently, Monte Carlo tree search (MCTS), a popular reinforcement learning algorithm \citep{mcts_survey} previously proven successful in a variety of domains like playing puzzle games such as Chess and Go \citep{mcts_alphago}, and was used by \citet{qroute_mcts} to develop a qubit routing solution. 


\subsection{\label{sec:intro-contribution}Our Contributions}

Our work demonstrates the use of MCTS on the task of Qubit Routing and presents state of the art results. Following are the novelties of this approach:

\begin{itemize}
    \item We use an array of mutex locks to represent the current state of parallelization, helping to reduce the depth of the circuits instead of the total quantum volume, in contrast to previous use of MCTS for qubit routing in \citet{qroute_mcts}.
    \item The actions in each timestep (layer of the output circuit) belong to a innumerably large action space. We phrase the construction of such actions as a Markov decision process, making the training stabler and the results better, particularly at larger circuit sizes, than those obtained by performing simulated annealing to search in such action spaces \cite{qroute_dqn1, qroute_dqn2}. Such approach should be applicable to other problems of a similar nature.
    \item Graph neural networks are used as an improved architecture to help guide the tree search.
\end{itemize}

Finally, we provide a simple python package containing the implementation of QRoute, together with  an easy interface for trying out different neural net architectures, combining algorithms, reward structures, etc.

\begin{figure*}[ht]
    \centering
    \includegraphics[width=0.75\linewidth]{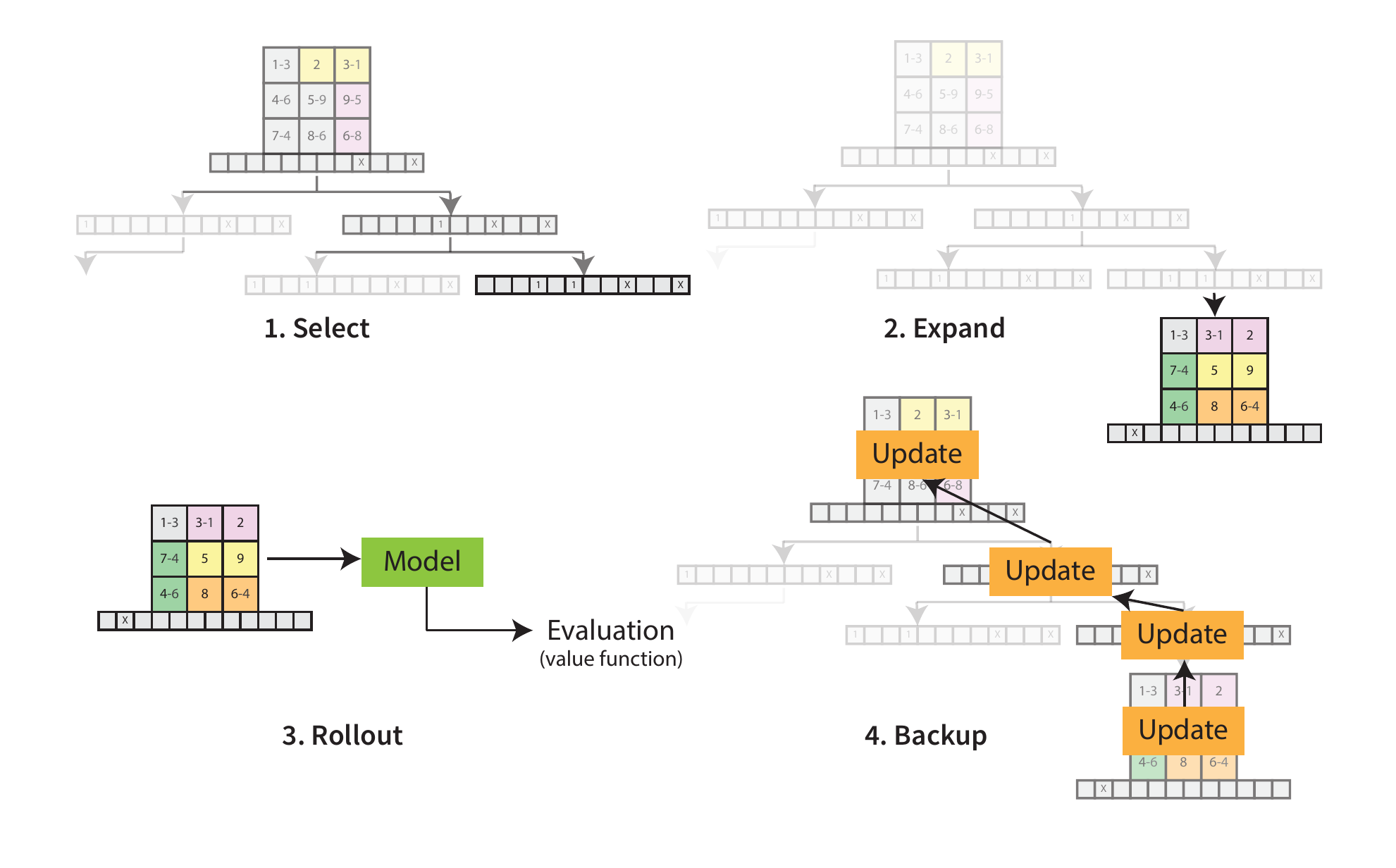}
    \caption{\label{fig:mcts-explainer}
        Iteration of a Monte Carlo tree search: (i) select - recursively choosing a node in the search tree for exploration using a selection criteria, (ii) expand - expanding the tree using the available actions if an unexplored leaf node is selected, (iii) rollout - estimating long term reward using a neural network for the action-state pair for the new node added to search tree, and (iv) backup - propagating reward value from the leaf nodes upwards to update the evaluations of its ancestors.}
\end{figure*}

\newtheorem{defn}{Definition}[section]

\section{\label{sec:method}Method}

The QRoute algorithm takes in an input circuit and an injective map, $\mathcal{M}: Q \rightarrow N$, from logical qubits to nodes (physical qubits). Iteratively, over multiple timesteps, it tries to schedule the gate operations that are present in the input circuit onto the target hardware. To do so, from the set of unscheduled gate operations, $\mathcal{P}$, it takes all the current operations, which are the first unscheduled operation for both the qubits that they act on, and tries to make them into local operations, which are those two-qubit operations that involve qubits that are mapped to nodes connected on the target hardware.

In every timestep $t$, QRoute starts by greedily scheduling all the operations that are both current and local in $\mathcal{P}$. To evolve $\mathcal{M}$, it then performs a Monte Carlo tree search (MCTS) to find an optimal set of SWAPs by the evaluation metrics described in the Section \ref{sec:method-mcts} such that all operations in the current timestep put together form a parallelizable set, i.e., a set of local operations such that no two operations in the set act on the same qubit. The number of states we can encounter in the action space explodes exponentially with the depth of our search, therefore an explicit search till the circuit is done compiling is not possible. Therefore we cut short our search at some shallow intermediate state, and use a neural network to get its heuristic evaluation.

The following subsections describe in greater detail the working of the search and the heuristic evaluation.

\subsection{\label{sec:method-state} State and Action Space}

\begin{defn}[State]
    It captures entire specification of the state of compilation at some timestep t. Abstractly, it is described as:
    \begin{equation}
        s_{t} = (\mathcal{D}, \ \mathcal{M}_{t},\ \mathcal{P}_{t},\ \mathcal{L}_{t})
    \end{equation}

    where, $\mathcal{D}$ is the topology of target hardware, and $\mathcal{M}_t$ and $\mathcal{P}_t$ represents the current values of $\mathcal{M}$ and $\mathcal{P}$ respectively. $\mathcal{L}_{t}$ is the set of nodes that are locked by the gate operations from the previous timestep and therefore cannot be operated in the current timestep. 
\end{defn}

\begin{defn}[Action]
    It is a set of SWAP gates (represented by the pair of qubits it acts on) such that all gates are local, and its union with the set of operations that were scheduled in the same timestep forms a parallelizable set.
\end{defn}

We are performing a tree search over state-action pairs. Since the number of actions that can be taken at any timestep is exponential in the number of connections on the hardware, we are forced to build a single action up, step-by-step. 

\begin{defn}[Move]
    It is a single step in a search procedure which either builds up the action or applies it to the current state. Moves are of the following two types:
    \begin{enumerate}
        \item SWAP($n_1$, $n_2$): Inserts a new SWAP on nodes $n_1$ and $n_2$ into the action set. Such an insertion is only possible if the operation is local and resulting set of operations for the timestep form a parallel set.
        \item COMMIT: Finishes the construction of the action set for that timestep. It also uses the action formed until now to update the state $s_t$ (schedules the SWAP gates on the hardware), and resets the action set for the next step.
    \end{enumerate}
\end{defn}

In reality, different gate operations take different counts of timesteps for execution. For example, if a hardware requires SWAP gate to be broken down into CNOT gates, then it would take three timesteps for complete execution \citep{utk_equiv_circuits}. This means, operations which are being scheduled must maintain mutual exclusivity with other other operations over the nodes which participates in them. This is essential to minimizing the depth of the circuit since it models paralleizability of operations.

However, constructing a parallelizable set and representing the state of parallelization to our heuristic evaluator is a challenge. But an analogy can be drawn here to the nodes being thought of as ``resources" that cannot be shared, and the operations as ``consumers" \citep{mutex_dijkstra}. This motivates us to propose the use of Mutex Locks for this purpose. These will lock a node until a scheduled gate operation involving that node executes completely. Therefore, this allows our framework to naturally handle different types of operations which take different amounts of time to complete.


For every state-action pair, the application of a feasible move $m$ on it will result in a new state-action pair: $(s,a) \xrightarrow[]{m} (s^\prime,a^\prime)$. This is a formulation of the problem of search as a Markov Decision Process. Associated with each such state-action-move tuple $((s, a), m)$, we maintain two additional values that are used by MCTS:

\begin{enumerate}
\item \textit{N-value} - The number of times we have taken the said move $m$ from said state-action pair $(s,a)$.
\item \textit{Q-value} - Given a reward function $\mathcal{R}$, it is the average long-term reward expected after taking said move $m$ over all iterations of the search. (Future rewards are discounted by a factor $\gamma$)

\begin{equation}
\begin{split}
    Q((s,a), m) =&\ \mathcal{R}((s,a), m)\ + \\ & \gamma \frac{\sum_{m^\prime} N((s^\prime, a^\prime), m^\prime) \cdot Q((s^\prime, a^\prime), m^\prime)}{\sum_{m^\prime} N((s^\prime, a^\prime), m^\prime)}
\end{split}
\end{equation}

\end{enumerate}

\begin{figure*}[th]
\centering
    \includegraphics[width=.72\textwidth]{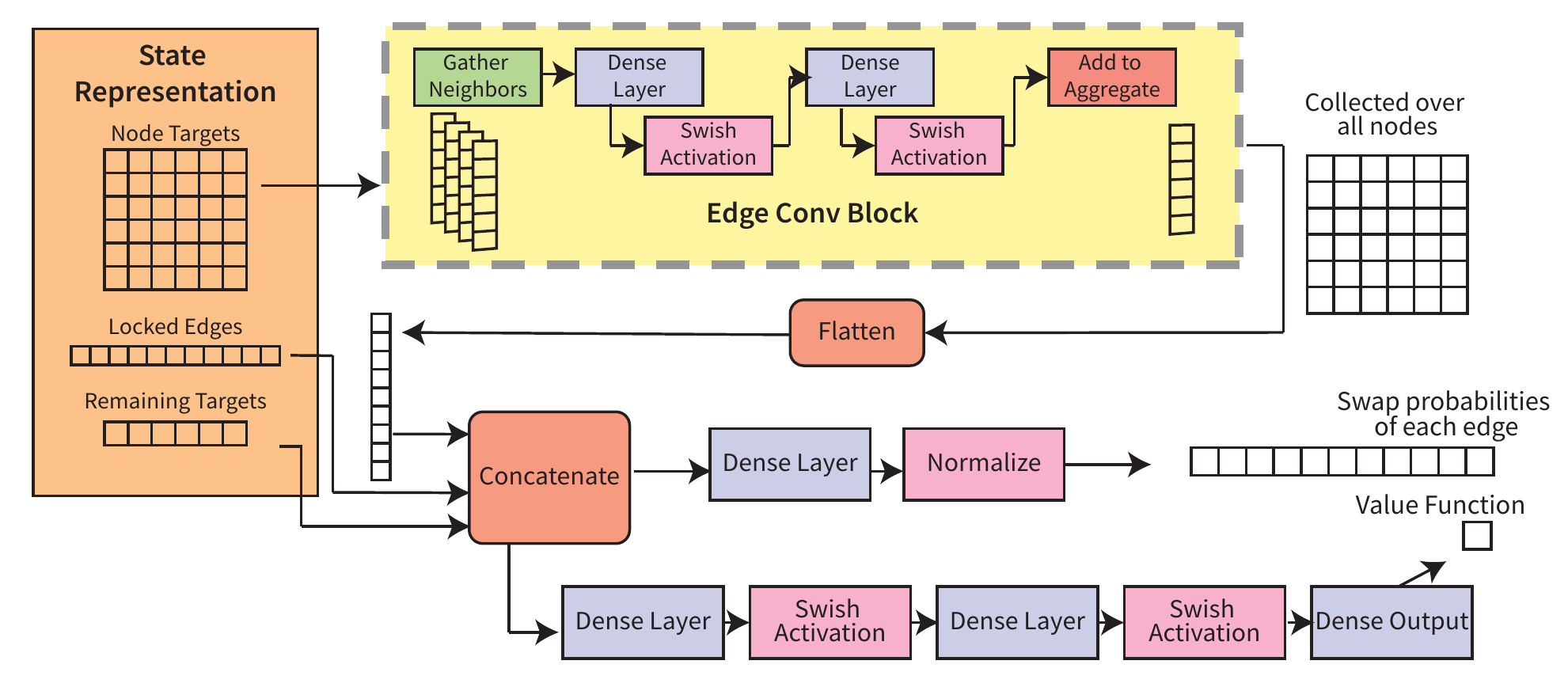}
    \caption{\label{fig:network-architecture}
     Graph neural network architecture that approximates the value function and the policy function.}
\end{figure*}

\subsection{\label{sec:method-mcts} Monte Carlo Tree Search}

Monte Carlo tree search progresses iteratively by executing its four phases: select, expand, rollout, and backup as illustrated in Fig. \ref{fig:mcts-explainer}. In each iteration, it begins traversing down the existing search tree by selecting the node with the maximum UCT value (Eq. \ref{eq:uct}) at each level. During this traversal, whenever it encounters a leaf node, it expands the tree by choosing a move $m$ from that leaf node. Then, it estimates the scalar evaluation for the new state-action pair and backpropagates it up the tree to update evaluations of its ancestors.

To build an optimal action set, we would want to select the move $m$ with the maximum true Q-value. But since true Q-values are intractably expensive to compute, we can only approximate the Q-values through efficient exploration. We use the Upper Confidence Bound on Trees (UCT) objective \citep{mcts_uct} to balance exploration and exploitation as we traverse through the search tree. Moreover, as this problem results in a highly asymmetric tree, since some move block a lot of other moves, while others block fewer moves, we use the formulation of UCT adapted for asymmetric trees \citep{mcts_assymetric}:

\begin{equation}\label{eq:uct}
\begin{split}
    \textrm{UCT}((s,a), m) =&\ Q((s,a), m)\ + \\ & c \frac{\sqrt{\sum_m N((s,a), m)}}{N((s,a), m)} \times p(m \vert (s,a))
\end{split}
\end{equation}

Here, the value $p(m | (s,a))$ is the prior policy function, which is obtained by adding a Dirichlet noise to the policy output of the neural network \citep{mcts_alphazero}. As MCTS continues probing the action space, it gets a better estimate of the true values of the actions. This means that it acts as a policy enhancement function whose output policy (Eq. \ref{eq:mcts_output}) can be used to train the neural network's prior ($\pi$), and the average Q-value computed can be used to train its scalar evaluation (Eq. \ref{eq:train_nn}).

\begin{equation}\label{eq:mcts_output}
    \pi(m | (s,a)) \propto N((s, a), m))
\end{equation}

\begin{equation}\label{eq:train_nn}
    \mathcal{V}((s,a)) = \frac{\sum_{m} Q((s,a), m)}{\sum_{m} N((s,a), m)}
\end{equation}

The details of how MCTS progresses have been elaborated in the supplementary. Once it gets terminated, i.e., the search gets completed, we go down the tree selecting the child with the maximum Q-value at each step until a COMMIT action is found, we use the action set of the selected state-action pair to schedule SWAPs for the current timestep, and we re-root the tree at the child node of the COMMIT action to prepare for the next timestep.

\subsection{\label{sec:method-model}Neural Network Architecture}

Each iteration of the MCTS requires evaluation of Q-values for a newly encountered state-action pair. But these values are impossible to be computed exactly since it would involve an intractable number of iterations in exploring and expanding the complete search tree. Therefore, it is favorable to heuristically evaluate the expected long-term reward from the state-action pair using a Neural Network, as it acts as an excellent function approximator that can learn the symmetries and general rules inherent to the system.

So, once the MCTS sends a state-action pair to the evaluator, it begins by committing the action to the state and getting the resultant state. We then generate the following featurized representation of this state and pass this representation through the neural-network architecture as shown in Fig. \ref{fig:network-architecture}.

\begin{enumerate}
    \item \textit{Node Targets} - It is a square boolean matrix whose rows and columns correspond to the nodes on a target device. An element $(i, j)$ is true iff some logical qubits $q_x$ and $q_y$ are currently mapped to nodes $i$ and $j$ respectively, such that ($q_x$, $q_y$) is the first unscheduled operation that $q_x$ partakes in.
    \item \textit{Locked Edges} - It is a set of edges (pairs of connected nodes) that are still locked due to either of its qubits being involved in an operation in the current timestep or another longer operation that hasn't yet terminated from the previous timesteps.
    \item \textit{Remaining targets} - It is a list of the number of gate-operations that are yet to be scheduled for each logical qubit. 
\end{enumerate}


The SWAP operations each qubit would partake in depends primarily on its target node, and on those of the nodes in its neighborhood that might be competing for the same resources. It seems reasonable that we can use a Graph Neural Network with the device topology graph for its connectivity since the decision of the optimal SWAP action for some node is largely affected by other nodes in its physical neighborhood. Therefore, our architecture includes an edge-convolution block \citep{nn_edge_conv}, followed by some fully-connected layers with Swish \citep{nn_swish} activations for the policy and value heads. The value function and the policy function computed from this neural network are returned back to the MCTS.

\section{\label{sec:results}Results}
We compare QRoute against the routing algorithms from other state-of-the-art frameworks on various circuit benchmarks: (i) Qiskit and its three variants \citep{comp_qiskit}: (a) basic, (b) stochastic, and (c) sabre, (ii) Deep-Q-Networks (DQN) from \citep{qroute_dqn2}, (iii) Cirq \citep{comp_cirq}, and (iv) t$\ket{\text{ket}}$ from Cambridge Quantum Computing (CQC) \citep{comp_pytket}. 
Qiskit's transpiler uses gate commutation rules while perform qubit routing. This strategy is shown to be advantageous in achieving lower circuit depths \citep{bridge_gate} but was disabled in our simulations to have a fair comparison. The results for DQN shown are adapted from the data provided by the authors \citet{qroute_dqn2}.

\begin{figure}[t]
    \includegraphics[width=\linewidth]{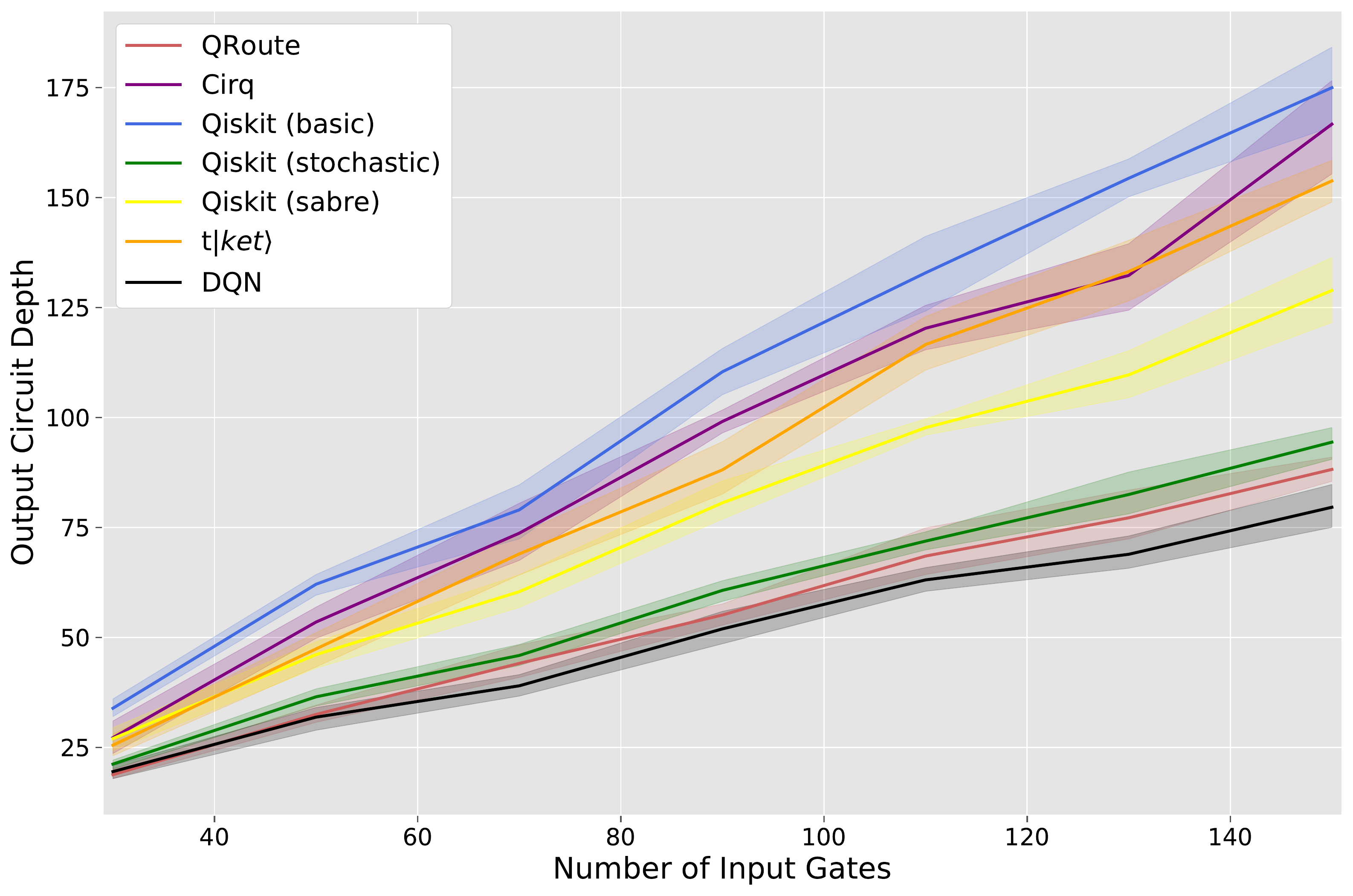}
    \caption{\label{fig:results-random}
        Comparative performance of routing algorithms on random circuits as a function of the number of two-qubit operations in the circuit.}
\end{figure}

\subsection{\label{sec:results-random}Random Test Circuits}

The first benchmark for comparing our performance comprises of random circuits. These circuits are generated on the fly and initialized with the same number of qubits as there are nodes on the device. Then two-qubit gates are put between any pair of qubits chosen at random. In our simulations, the number of such gates is varied from 30 to 150 and the results for assessing performance of different frameworks are given in Fig. \ref{fig:results-random}. The experiments were repeated 10 times on each circuit size, and final results were aggregated over this repetition.

Amongst the frameworks compared, QRoute ranks a very close second only to Deep-Q-Network guided simulated annealer (DQN). Nevertheless, QRoute still does consistently better than all the other major frameworks: Qiskit, Cirq and t$\ket{\text{ket}}$, and it scales well when we increase the number of layers and the layer density in the input circuit. QRoute shows equivalent performance to DQN on smaller circuits, and on the larger circuits it outputs depths which are on average $\leq 4$ layers more than those of DQN. Some part of this can be attributed to MCTS, in it's limited depth search, choosing the worse of two moves with very close Q-values, resulting in the scheduling of some unnecessary SWAP operations.

\subsection{\label{sec:results-small}Small Realistic Circuits}

\begin{figure}[t]
    \includegraphics[width=\linewidth]{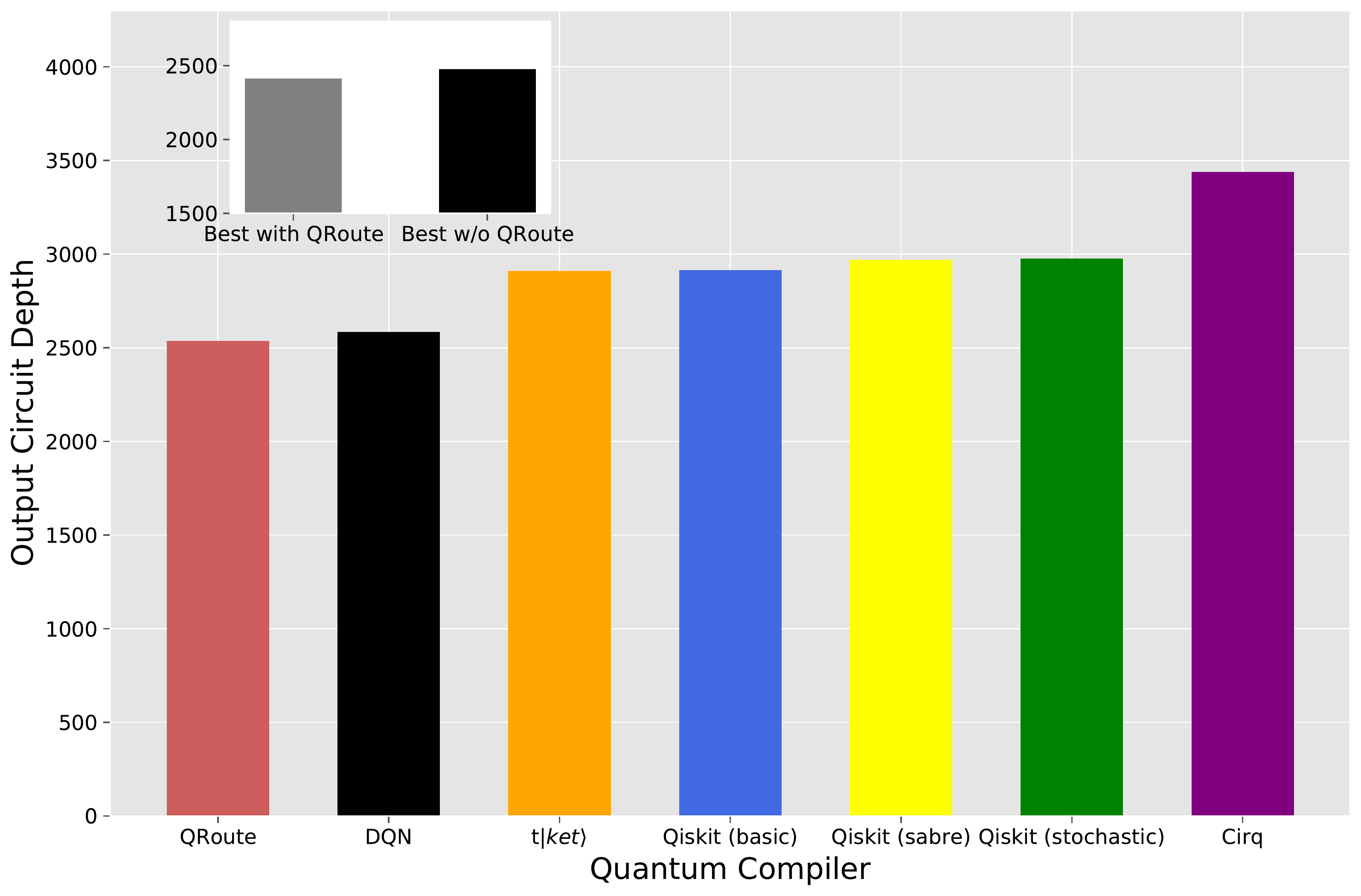}
    \caption{\label{fig:results-small}
        Plots of output circuit depths of routing algorithms over small realistic circuits ($\leq 100$ gates), summed over the entire dataset. The inset shows the results on the same data comparing the best performant scheduler excluding and including QRoute on each circuit respectively.}
\end{figure}

Next we test on the set of all circuits which use 100 or less gates from the IBM-Q realistic quantum circuit dataset used by \citet{data_realistic}. The comparative performance of all routing frameworks has been shown by plotting the depths of the output circuits summed over all the circuits in the test set in Fig. \ref{fig:results-small}. Since the lack of a good initial qubit allocation becomes a significant problem for all pure routing algorithms on small circuits, we have benchmarked QRoute on this dataset from three trials with different initial allocations.

The model presented herein has the best performance on this dataset. We also compare the best result from a pool of all routers including QRoute against that of another pool of the same routers but excluding QRoute. The pool including QRoute gives on average $2.5\%$ lower circuit depth, indicating that there is a significant number of circuits where QRoute is the best routing method available.

\begin{figure*}[ht]
    \centering
    \includegraphics[width=.84\linewidth]{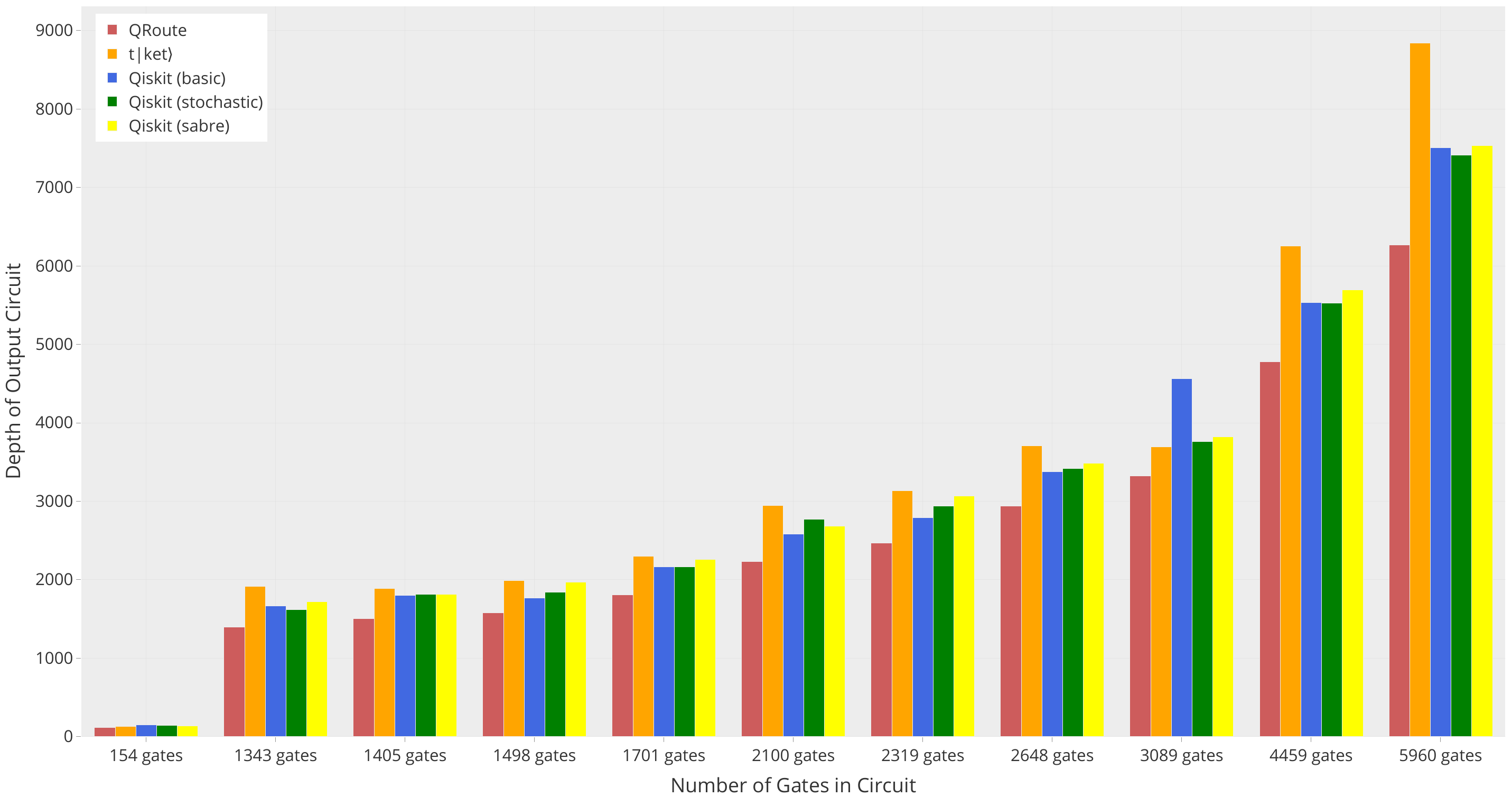}
    \caption{\label{fig:results-large}
        The results over eight circuits sampled from the large realistic dataset benchmark, the outputs of each routing algorithm are shown for every circuit.}
\end{figure*}

On this dataset also, closest to QRoute performance is shown by Deep-Q-Network guided simulated annealer. To compare performances, we look at the average circuit depth ratio (CDR), which is defined by \citep{qroute_dqn2}:
\begin{equation}
    \text{CDR} = \frac{1}{\textrm{\#circuits}} \sum_{\textrm{circuits}} \frac{\textrm{Output Circuit Depth}}{\textrm{Input Circuit Depth}}
    \label{eq:CDR}
\end{equation}
The resultant CDR for QRoute is $1.178$, where as the reported CDR for the DQN is $1.19$. In fact, QRoute outperforms DQN on at least $80\%$ of the circuits. This is significant because in contrast to the random circuit benchmark, the realistic benchmarks consist of the circuits that are closer to the circuits used in useful computation.

\subsection{\label{sec:results-realistic}Large Realistic Circuit}

For final benchmark, we take eight large circuits ranging from 154 gates to 5960 gates in its input from the IBM-Quantum realistic test dataset \citep{data_realistic}. The results are plotted in Fig.  \ref{fig:results-large}. QRoute has the best performance of all available routing methods: Qiskit and t$\ket{\text{ket}}$, on every one of these sampled circuits with on an average $13.6\%$ lower circuit depth, and notable increase in winning difference on the larger circuits.

The results from DQN and Cirq are not available for these benchmarks as they are not designed to scale to such huge circuits. In case of DQN, the CDR data results were not provided for the circuits over $200$ gates, mainly because simulated annealing used in it is computationally expensive. Similarly, for Cirq, it takes several days to compile each of the near $5000$ qubit circuits. In contrast, QRoute is able to compile these circuits in at most 4 hours, and its compilation process can be sped up by reducing the depth of the search. Spending more time, however, helps MCTS to better approximate the Q-values leading to circuits with lower resulting depth.

\section{\label{sec:discussion-conclusion}Discussion and Conclusion}

In this article, we have shown that the problem of qubit routing has a very powerful and elegant formulation in Reinforcement Learning (RL) which can surpass the results of any classical heuristic algorithm across all sizes of circuits and types of architectures. Furthermore, the central idea of building up solutions step-by-step when searching in combinatorial action spaces and enforcing constraints using mutex locks, can be adapted for several other combinatorial optimization problems \citep{comb_survey, comb_1, comb_2, comb_3, comb_4}. Our approach is flexible enough to compile circuits of any size onto any device, from small ones like IBMQX20 with 20 qubits, to much larger hardware like Google Sycamore (results provided in supplementary) with 53 qubits (the Circuit Depth Ratio for small realistic circuits on Google Sycamore was 1.64). Also, it intrinsically deals with hardware having different primitive instruction set, for example on hardware where SWAP gates are not a primitive and they get decomposed to 3 operations. QRoute enjoys significant tunability; hyperparameters can be changed easily to alter the tradeoff between time taken and optimality of decisions, exploration and exploitation, etc.

QRoute is a reasonably fast method, taking well under 10 minutes to route a circuit with under 100 operations, and at most 4 hours for those with upto 5000 operations, when tested on a personal machine with an i3 processor (3.7 GHz) and no GPU acceleration. Yet more can be desired in terms of speed. However, it is hard to achieve any significant improvement without reducing the number of search iterations and trading off a bit of performance. More predictive neural networks can help squeeze in better speeds.

One of the challenges of methods like DQN, that use Simulated Annealing to build up their actions is that the algorithm cannot plan for the gates which are not yet waiting to be scheduled, those which will come to the head of the list once the gates which are currently waiting are executed \citep{qroute_dqn2}. QRoute also shares this deficiency, but the effect of this issue is mitigated by the explicit tree search which takes into account the rewards that will be accrued in the longer-term future. There is scope to further improve this by feeding the entire list of future targets directly into our neural network by using transformer encoders to handle the arbitrary length sequence data. This and other aspects of neural network design will be a primary facet of future explorations. Another means of improving the performance  would be to introduce new actions by incorporating use of BRIDGE gates \citep{bridge_gate} and gate commutation rules \citep{utk_equiv_circuits} alongside currently used SWAP gates. The advantage of former is that it allows running CNOT gates on non-adjacent qubit without permuting the ordering of the logical qubits; whereas, the latter would allow MCTS to recognize the redundancy in action space, making its exploration and selection more efficient.

Finally, we provide an open-sourced access to our software library. It will allow researchers and developers to implement variants of our methods with minimal effort. We hope that this will aid future research in quantum circuit transformations. For review we are providing, the codebase and a multimedia in the supplementary.  

On the whole, the Monte Carlo Tree Search for building up solutions in combinatorial action spaces has exceeded the current state of art methods that perform qubit routing. Despite its success, we note that QRoute is a primitive implementation of our ideas, and there is great scope of improvement in future. 

\section*{Acknowledgements}
A. S. and U. A. have contributed equally to the
work presented in this manuscript. A. S. and U. A. would like to acknowledge the assistance from their colleague Bhuvanesh Sridharan for his help in implementing and refining MCTS. Additionally, they would also like to thank their colleagues Jai Bardhan, Kalp Shah for their suggestions.

\bibliography{qroute}

\end{document}